\documentclass[prd,twocolumn,showpacs,preprintnumbers,amsmath,amssymb,nofootinbib,APS]{revtex4-1}

\usepackage{dcolumn}
\usepackage{bm}
\usepackage[latin1]{inputenc}
\usepackage[spanish,english]{babel}
\usepackage{amsfonts}
\usepackage{amssymb}
\usepackage{graphicx}

\newcommand{\be}{\begin{equation}}
\newcommand{\ee}{\end{equation}}
\newcommand{\bea}{\begin{eqnarray}}
\newcommand{\eea}{\end{eqnarray}}

\begin{document}
\title{{\bf Adiabatic regularization for  spin-$1/2$ fields }}

\author{Aitor Landete, José Navarro-Salas and Francisco Torrentí}
\affiliation{ {\footnotesize Departamento de Fisica Teórica and
IFIC, Centro Mixto Universidad de Valencia-CSIC, Facultad de Física, Universidad de Valencia,
        Burjassot-46100, Valencia, Spain. \ \ }}

\date{May 31, 2013}

\begin{abstract}

We  extend  the adiabatic regularization method  to spin-$1/2$ fields. The ansatz for the adiabatic expansion for fermionic modes differs significantly from the WKB-type template that works for scalar modes.
 We give explicit expressions for the first adiabatic orders  and analyze particle creation in de Sitter spacetime.  As for scalar fields, the adiabatic method can be distinguished by its capability to overcome the UV divergences of the particle number operator.
We also test  the consistency of the extended method  by working out the conformal and axial anomalies for a Dirac field in a  FLRW spacetime, 
in exact agreement with those
 obtained from other renormalization prescriptions. We finally show its power by computing the  renormalized stress-energy tensor  for Dirac fermions in de Sitter space.

\end{abstract}


\maketitle


{\it Introduction.} 
Quantum field theory in curved spacetime offers a first step to merge Einstein's theory of general relativity and quantum field theory in Minkowski space within a self-consistent and successful framework \cite{parker-toms, birrell-davies}. The  discovery of particle creation in a time-dependent gravitational field \cite{parker66, parker69} has proved  of paramount importance.  It constitutes the driving mechanism to explain the quantum radiance in a  gravitational collapse producing a black hole \cite{hawking74} 
and the generation of  cosmic primordial inhomogeneities, observed now in the cosmic microwave background and the large-scale structure of the Universe \cite{Liddle-Lyth00}.  The gravitationally created particles   generate an energy density with new ultraviolet (UV) divergences, as compared with the UV divergences present in Minkowski space. This requires more sophisticated methods of renormalization, adapted to the time-dependent or curved background.  

Adiabatic regularization was first introduced in Parker's pioneer work on particle creation in the expanding universe  \cite{parker66} as a way to overcome the rapid oscillation of the particle number operator and UV divergences during the expansion. 
The method was  later systematized and generalized  \cite{parker-fulling74} to consistently deal with the UV divergences of the  stress-energy tensor of scalar fields. 
The adiabatic method  identifies  the UV subtraction terms by first considering a slowly varying expansion factor $a(t)$. This naturally leads to 
a Liouville or WKB-type asymptotic expansion for the modes characterized by the comoving momentum $\vec{k}$. The subtraction terms identified this way are valid for arbitrary smooth expansions. The method was originally designed to deal with the particle number operator and it is a distinguishing feature of adiabatic renormalization. When the method is applied to renormalize local expectation values, as the stress-energy tensor, it turns out to be equivalent to the DeWitt-Schwinger point-splitting prescription for scalar fields \cite{Birrell78, anderson-parker}. An advantage of  adiabatic regularization is that it is very efficient for numerical calculations \cite{Hu-Parker, Anderson, Anderson2}. It  is  also potentially important to scrutinize the power spectrum in inflationary cosmology \cite{parker07}. It  also plays a crucial  role in the understanding of the low-energy regime in quantum cosmology \cite{agullo-ashtekar-nelson}.

The point-splitting prescription \cite{dewitt75, christensen76} 
can be naturally extended  to  spin-$1/2$  fields \cite{christensen78}, and one would expect an analogous extension within the adiabatic subtraction scheme. However, a systematic adiabatic expansion for spin one-half modes, required to identify the subtraction terms,  has  been elusive.  In this paper we provide a basis for such  expansion and prove it by working out the axial vector current  and the conformal anomalies in a Friedmann-Lemaître-Robertson-Walker (FLRW) universe,  and also analyzing particle creation and the renormalized stress-energy tensor in de Sitter spacetime. 

We have to remark that  the existence of a well-defined extension of adiabatic regularization for spin-$1/2$ fields can be expected on physical grounds. 
As stressed before, and first showed in the seminal work on particle creation \cite{parker66}, the adiabatic scheme can be distinguished from other renormalization methods because it is a unique  method to  overcome the UV divergences that appear 
in the particle number operator.  One would expect that both the mean particle number  and its uncertainty would be  well-defined for fermions in a slowly expanding universe. Only a consistent adiabatic method for spin-$1/2$ fields enforces this physical requirement.

{\it Adiabatic regularization for scalar fields.} 
A  scalar field $\phi$ satisfying the  wave equation $ (\Box + m^2 +\xi R) \phi =0$ 
 can be expanded (for simplicity we assume a  spatially flat FLRW  universe  $ds^2=dt^2- a^2(t)d\vec{x}^2$)  in the form $\phi= \sum_{\vec{k}} ( A_{\vec{k}}f_{\vec{k}}(\vec{x}, t) +  A^{\dagger}_{\vec{k}}f^*_{\vec{k}}(\vec{x}, t))$, 
where the modes   are $ f_{\vec{k}}=(2L^3a^3(t))^{-1/2}e^{i\vec{k}\vec{x}}h_k(t)$ ($k=|\vec{k}|$).
For convenience, we have assumed periodic boundary conditions in a cube of comoving length $L$. Therefore, $k^i= 2\pi n^i/L$ with $n^i$ an integer. Later on we shall take the continuous limit $L \to \infty$.
These modes are forced to obey the normalization condition with respect to the conserved Klein-Gordon product $(f_{\vec{k}}, f_{\vec{k}'})= \delta_{\vec{k},\vec{k}'}$. 
This condition translates to a Wronskian-type condition for the functions $h_k(t)$:  $ h_k^*\dot h_k - \dot h_k^{*} h_k = -2i $
 (the dot means derivative with respect to proper time $t$).
We also have that $(f_{\vec{k}}, f^*_{\vec{k}'})= 0$.
These conditions ensure the basic commutation relations for annihilation and creation operators.
Adiabatic regularization is based on a generalized WKB-type asymptotic expansion of the modes according to the ansatz \cite{parker-toms}
$h_k(t) = \frac{1}{\sqrt{W_k(t)}}e^{-i\int^t W_k(t')dt'}$.
Note that this ansatz guarantees automatically the  Wronskian condition.
The equation for $h_k$ reads $\ddot h_k + (\omega_k^2 + \sigma)h_k=0$,
where $\omega_k(t)= \sqrt{k^2/a^2(t) + m^2}$ and $\sigma = (6\xi -3/4)\dot a^2/a^2 + (6\xi -3/2)\ddot a/a$. It 
translates to the following equation for the function $W_k(t)$: $W_k^2 = \omega_k^2 + \sigma + W_k^{-1/2}\frac{d^2}{dt^2}W_k^{-1/2}$.
One then expands $W_k$ in an adiabatic series, determined by the number of time derivatives of the expansion factor $a(t)$: $W_k(t) = \omega^{(0)}(t) + \omega^{(2)}(t)+ \omega^{(4 )}(t) + ... $,
where the leading term $\omega^{(0)}(t)\equiv \omega(t)\equiv \omega_k(t)= \sqrt{k^2/a^2(t) + m^2}$ is the usual redshifted frequency.  The higher adiabatic terms are obtained by iteration. The second order adiabatic contribution, which depends on $\ddot a$ and $\dot a^2$ is 
$\omega^{(2)}(t)= \frac{\sigma }{2\omega} + \frac{1}{2}\omega^{-1/2}\frac{d^2}{dt^2}\omega^{-1/2}$.
The iteration can be applied indefinitely to get  any $\omega^{(n)}$.
For a  slowly  varying $a(t)$ the above series expansion allows to define the particle number as an adiabatic invariant \cite{parker66, parker69}. Furthermore, the  UV divergences 
of the variance and the stress-energy tensor can be removed by subtraction of the corresponding contributions, mode by mode, to second and fourth adiabatic order, respectively \cite{parker-fulling74}.  This procedure of removing the UV divergences preserves covariance and leads to finite expectation values for the stress-energy tensor that obey  covariant conservation. 
After this brief introduction on the adiabatic method for scalar fields we present now our proposal for extending it to spin-$1/2$ fields.\\

 {\it Adiabatic expansion for spin-$1/2$ fields.}
Let us consider the Dirac equation in  a spatially flat  FLRW spacetime 
\bea &(i\gamma^0\partial_0+ \frac{3i}{2}\frac{\dot a }{a}\gamma^0  +\frac{i}{a}\vec{\gamma}\vec{\nabla} -m)\psi=0 \ , \eea
where $\gamma^{\mu}$ are the  Dirac matrices in Minkowski spacetime.
 For our purposes it is convenient to work with  the standard  Dirac-Pauli representation. 
After  momentum expansion $ \psi=\sum_{\vec{k}} \psi_{\vec{k}}(t)e^{i\vec{k}\vec{x}}$ it is convenient
to write the Dirac field in terms of  two two-component spinors
\bea \label{psik}
&\psi_{\vec{k}}(t)=
\left( {\begin{array}{c}
  \frac{1}{\sqrt{L^3 a^3}}h^I_{{k}}(t) \xi_{\lambda} (\vec{k}) \\
  \frac{1}{\sqrt{L^3 a^3}}h^{II}_{{k}}(t)\frac{\vec{\sigma}\vec{k}}{k} \xi_{\lambda} (\vec{k})\\
 \end{array} } \right)
\eea
where $\vec{\sigma}$ are the usual Pauli matrices. 
$ \xi_{\lambda} (\vec{k})$ is a constant  normalized two-component spinor $\xi_{\lambda}^{\dagger}\xi_{\lambda}=1$ such that  
 $\frac{\vec{\sigma}\vec{k}}{2k}\xi_{\lambda}= \lambda \xi_{\lambda}$.
$\lambda ={\pm} 1/2$  represents the eigenvalue for the helicity, or spin component along the $\vec{k}$ direction.
$h_{{k}}^I$ and $h_{{k}}^{II}$ are scalar functions, obeying the coupled first order equations   
\bea &h_{{k}}^{II}=\frac{ia}{k}(\partial_t+im)h_{{k}}^I  \ \ , \ \  h_{{k}}^{I}=\frac{ia}{k}(\partial_t-im)h_{{k}}^{II} \label{eq:hII} , \eea
and the uncoupled second order equations:
$(\partial_t^2 +\frac{\dot a}{a}\partial_t  +im\frac{\dot a}{a}+m^2 + \frac{k^2}{a^2} )h_{{k}}^I=0$ and $(\partial_t^2 +\frac{\dot a}{a}\partial_t  -im\frac{\dot a}{a}+m^2 + \frac{k^2}{a^2} )h_{{k}}^{II}=0$. 
The normalization condition for the four-spinor is 
\bea \label{normalization} &|h_{{k}}^I(t)|^2 +   |h_{{k}}^{II}(t)|^2 =1 \ . \eea
This condition guaranties the standard anticommutator relations for creation and annihilation operators  defined by the expansion
$ \psi= \sum_{\vec{k}} \sum_{\lambda = {\pm 1/2}} ( B_{\vec{k}, \lambda } u_{\vec{k}, \lambda}(t, \vec{x}) +  D^{\dagger}_{\vec{k}, \lambda} v_{\vec{k}, \lambda} (t, \vec{x}) )$,
where $u_{\vec{k}, \lambda}(t, \vec{x})$ is defined from an exact solution to the above equations. 
The orthogonal modes $ v_{\vec{k}, \lambda} (t, \vec{x})$ are obtained by the charge conjugation operation $v_{\vec{k}, \lambda}=Cu_{\vec{k}, \lambda}=i\gamma^2u^*_{\vec{k}, \lambda}$. 
One could be tempted to use the above second order equations to generate a WKB-type expansion for $h_k^I$ and $h_k^{II}$. 
However, the WKB  ansatz  is specifically designed to preserve the Klein-Gordon product, and hence the associated Wronskian condition, but not to preserve the Dirac product and the  (normalization) condition (\ref{normalization}). 
Therefore, one should follow a different route. (For a study of fermion pair production in Minkowski space using the WKB ansatz see \cite{Kluger92}). 

The zeroth adiabatic order should naturally generalize  the standard solution in Minkowski space.
Therefore, it must be of the form 
\bea &g^{I (0)}_{{k}}(t) = \sqrt{\frac{\omega (t) + m}{2\omega(t)}} e^{-i\int^t \omega (t') dt'} \nonumber \\
&g^{II(0)}_{{k}}(t) = \sqrt{\frac{\omega(t) - m}{2\omega(t)}} e^{-i\int^t\omega (t') dt'} \ . \eea
 It is easy to see that the zeroth order obeys the normalization condition $|g_{{k}}^{I (0)}(t)|^2 +   |g_{{k}}^{II(0)}(t)|^2 =1$. The form of the zeroth order and the field equations  (\ref{eq:hII}) 
 suggests the following alternative ansatz for the  adiabatic expansion (at order $n$)
\bea   \label{nadiabatic} &g^{I (n)}_{{k}}(t) = \sqrt{\frac{\omega  + m}{2\omega}} e^{-i\int^t (\omega (t') + \omega^{(1)}+ ... +\omega^{(n)})dt'} \nonumber \\ &\times (1+ F^{(1)}+... +F^{(n)}) \nonumber \\
 &g^{II (n)}_{{k}}(t) = \sqrt{\frac{\omega - m}{2\omega}} e^{-i\int^t(\omega (t') + \omega^{(1)}+ ... +\omega^{(n)})dt'} \nonumber \\ &\times (1+G^{(1)}+  ... + G^{(n)}) \ ,   \eea
where $\omega^{(n)}$, $F^{(n)}$, and $G^{(n)}$ are local functions of adiabatic order $n$. 
Imposing Eqs. (\ref{eq:hII}) 
and keeping terms of fixed adiabatic order, one gets a system of equations at each order. Moreover, the solution should also 
respect the normalization condition $|g_{{k}}^{I (n) }(t)|^2 +   |g_{{k}}^{II (n)}(t)|^2 =1$ (at the given adiabatic order $n$), which we impose as a new equation. For the adiabatic order one we obtain immediately that $\omega^{(1)}=0$. Moreover, the functions $F^{(1)}$, $G^{(1)}$ should have a vanishing real part and verify the single relation $G^{(1)}  =  F^{(1)} +i\frac{m \dot a}{ 2\omega^2 a}$.  The solution can be parametrized as $F^{(1)}=-A i\frac{m \dot a}{ \omega^2 a}$, $G^{(1)}= B i\frac{m \dot a}{ \omega^2 a}$, where $A, B$ are arbitrary real constants obeying $A+B=1/2$.  We can go further and consider the system of equations at adiabatic order two. We note that, although  the solution at first order is not univocally determined, local observables are actually independent of the ambiguity in $A-B$.  We find it useful for simplifying  expressions and for computational purposes to fix the parameters  as $A=B$. This implies $F^{(1)}(-m)= G^{(1)}(m)$, $F^{(2)}(-m)= G^{(2)}(m)$, and so forth. The solutions are then (where $R= 6(\ddot a/a + \dot a^2/a^2)$)
\bea  &\omega^{(2)}= \frac{5m^4\dot a^2 -3\omega^2m^2\dot a^2 -2\omega^2m^2\ddot a a}{8\omega^5a^2}  \\
  &F^{(2)}= \frac{m^2R}{48\omega^4} -\frac{5m^4\dot a^2}{16\omega^6a^2}-\frac{m^2\dot a^2}{32\omega^4a^2} - \frac{mR}{48\omega^3} + \frac{5m^3\dot a^2}{16\omega^5 a^2} \nonumber \\
  &G^{(2)}= \frac{m^2R}{48\omega^4} -\frac{5m^4\dot a^2}{16\omega^6a^2}-\frac{m^2\dot a^2}{32\omega^4a^2} + \frac{mR}{48\omega^3} - \frac{5m^3\dot a^2}{16\omega^5 a^2} \nonumber \ . \eea
We can continue the iteration in a systematic way, where we find $\omega^{(odd)}=0$. The explicit solutions to third and fourth adiabatic orders   will be given elsewhere. 
The adiabatic $n$th order fermionic modes  defined by $g_k^{I(n)}$ and $g_k^{II(n)}$ allow us to define the subtraction terms to cancel the UV divergences. A first divergence appears in the analysis of the particle number of created particles during a generic expansion of the Universe. A second worry  concerns the covariance of the subtraction scheme when it is typically applied  to the renormalization of the stress-energy tensor. 
To show that our proposal is able to solve satisfactorily these challenges, we will consider two physically relevant questions: particle creation in de Sitter space and the  conformal anomaly  in a FLRW spacetime. \\

{\it Particle creation in de Sitter spacetime.}
We focus now on the application of the  adiabatic expansion for the particle creation process in de Sitter spacetime $a(t)=e^{Ht}$.
The exact modes defining the analogous state of the Bunch-Davies vacuum are
given by $h_k^{I}= iNe^{-Ht/2} H^{(1)}_{\nu}(z)$ and $
h_k^{II}= Ne^{-Ht/2}H^{(1)}_{\nu-1}(z)$, 
 with $z\equiv ke^{-Ht}/H$, $N\equiv  \frac{1}{2}\sqrt{\frac{\pi k}{H}}e^{\pi m/2H}$, and $\nu=\frac{1}{2}-i\frac{m}{H}$. These  functions behave at very early times $t \to -\infty$ as the zeroth order adiabatic ones $g_k^{I (0)}$, $g_k^{II (0)}$.
As for bosons, the quantized field $\psi$ can also be expanded in terms of the fermionic $n$-order adiabatic modes $g^{(n)}_{\vec{k}, \lambda}(\vec{x}, t)$ [$g^{(n)c}_{\vec{k}, \lambda} (t, \vec{x}) $ are the corresponding ones obtained by the charge conjugation operation $C$] \bea &\psi= \sum_{\vec{k}, \lambda}  ( b^{(n)}_{\vec{k}, \lambda }(t) g^{(n)}_{\vec{k}, \lambda}(t, \vec{x}) +  d^{(n)\dagger}_{\vec{k}, \lambda}(t) g^{(n) c}_{\vec{k}, \lambda} (t, \vec{x}) )\ , \eea
  where $g^{(n)}_{\vec{k}, \lambda}$ are obtained from  (\ref{psik}) by replacing $h_k^{I, II}$ by $g_k^{I, II (n)}$.
 The time-dependent operators  $b^{(n)}_{\vec{k}, \lambda }(t)$ [$d^{(n)}_{\vec{k}, \lambda }(t)$]  are related to the time-independent ones $B_{\vec{k}, \lambda }$ ($D_{\vec{k}, \lambda }$) by a Bogolubov transformation. The corresponding Bogolubov coefficients, at a given adiabatic order $n$, can be obtained  from the functions $h_k^I(t)$ and $h_k^{II}(t)$ of the exact  modes by solving the system of equations (for simplicity we restrict to $\lambda=1/2$; similar equations apply for the opposite helicity)
\bea \label{systemhdothfermions} &h^I_k(t) = \alpha^{(n)}_k(t) g_k^{I (n)}- \beta^{(n)}_k(t)g_k^{II (n) *} \nonumber \\  &h^{II}_k(t) = \alpha^{(n)}_k(t) g_k^{II (n) }+ \beta^{(n)}_k(t)g_k^{I (n) *}  \ . \eea 
The average number of created fermionic particles with momentum $\vec{k}$, and with the given helicity (we omit the helicity index), is 
$ \langle N^{(n)}_{\vec{k}} \rangle= \langle b^{(n)\dagger}_{\vec{k}}(t)b^{(n)}_{\vec{k}}(t) \rangle = |\beta^{(n)}_k(t)|^2 $.
 In adiabatic regularization one should resort to the minimum adiabatic order required to obtain a UV finite result. For the average number density of total created  particles $\frac{1}{L^3 a^3} \sum_{\vec{k}} \langle N^{(n)}_{\vec{k}}(t)\rangle$ the required order is zero, since $|\beta^{(0)}_k(t)| \sim  O(k^{-2})$, as $k \to \infty$.   
 However,  the zeroth order is not enough to have a finite result for the sum of fluctuations. For a spin-$1/2$ field 
  the  sum of uncertainties  $ \Delta N^{(n)}_{\vec{k}}(t)\equiv (\langle N^{(n) 2}_{\vec{k}}\rangle- \langle N^{(n)}_{\vec{k}}\rangle^2)^{1/2} $ over all  momenta  has a linear UV divergence ($\Delta N^{(n)}_{\vec{k}}(t)\sim |\beta_k^{(n)}|$) when computed at the zeroth adiabatic  order. The minimal adiabatic order necessary to cancel this UV divergence is $2$, since 
 $|\beta^{(2)}_k(t)| \sim  O(k^{-4})$, while $|\beta^{(1)}_k(t)| \sim  O(k^{-3})$. The same falloff behavior appears for a generic expansion factor $a(t)$. As stressed in the introduction, this shows the necessity of the adiabatic regularization to properly define the particle number concept, even when the expansion is very slow.\\ 
 
{\it Conformal and axial vector current anomalies.} 
Concerning local observables in a generic FLRW spacetime, the second adiabatic order is required to renormalize  $\langle \bar \psi \psi \rangle$ and $\langle \bar \psi \gamma^5\psi \rangle$, while the fourth order is the right one to renormalize the stress-energy tensor $\langle T^{\mu}_{\nu}\rangle$ and $\langle \nabla_{\mu}J^{\mu}_A \rangle$, where $J^{\mu}_A$ is the axial vector current.   
 The renormalized values $\langle \bar \psi \psi \rangle_{r}$ and $\langle \bar \psi \gamma^5\psi \rangle_{r}$ are obtained by subtracting from the formal divergent expression the corresponding second-order adiabatic terms (we take here the continuous limit)
\bea \label{barpsipsi}&\langle \bar \psi \psi \rangle_{r}= \frac{-2}{(2\pi)^3a^3}\int d^3k(|h_k^{I}|^2 - |h_k^{II}|^2- |g_k^{I (2)}|^2 + |g_k^{II (2)}|^2) \nonumber \\
 \label{barpsi5psi}&\langle \bar \psi \gamma^5\psi \rangle_{r}= \frac{-2}{(2\pi)^3a^3}\int d^3k(h_k^{I*} h_k^{II}- h_k^{II*}h_k^{I}- g_k^{I (2)*}g_k^{II (2)} \nonumber \\ &+ g_k^{II (2)*} g_k^{I (2)})\nonumber \ , \eea
where the functions $h_k^{I, II}$ characterize the quantum state.  By construction the above integrals are UV finite. 
Note in passing that these expressions also allow for an efficient numerical estimation when the modes for the quantum state are difficult to manage analytically.
  As an application of the method, and also  as a test of the consistency and power of the adiabatic expansion given in this paper, we now calculate the axial vector current and the conformal anomalies  for the Dirac field. From the classical Dirac equation one gets $T^{\mu}_{\mu} = m\bar \psi \psi$ and $ \nabla_{\mu}J^{\mu}_A = 2im\bar \psi \gamma^5 \psi$. Thus, formally we have  $\langle T^{\mu}_{\mu} \rangle_{r}=  m \langle \bar \psi \psi \rangle$ and $\langle \nabla_{\mu}J^{\mu}_A \rangle_r= 2im\langle\bar \psi \gamma^5 \psi \rangle$. However, here $\langle \bar \psi \psi \rangle$ and $\langle \bar \psi \gamma^5\psi \rangle$ should not be their physical renormalized values (at second adiabatic order), since the physical expectation values   $\langle T_{\mu}^{\ \nu} \rangle_r$ and $ \langle \nabla_{\mu}J^{\mu}_A \rangle_r$ are obtained by subtractions up to  the fourth adiabatic order. Therefore, we have to subtract in  $\langle \bar \psi \psi \rangle$ and $\langle \bar \psi \gamma^5\psi \rangle$ up to the fourth adiabatic order 
  \be\langle T^{\mu}_{\mu} \rangle_{r}= \frac{-2m}{(2\pi)^3a^3}\int d^3k(|h_k^{I}|^2 - |h_k^{II}|^2- |g_k^{I (4)}|^2 + |g_k^{II (4)}|^2)\ , \nonumber  \ee
and an analogous expression for the divergence of the axial vector current.
To evaluate the anomalies  we must take the limit $m \to 0$ at the end of the calculation. Concerning the axial current anomaly, the  subtraction terms of fourth adiabatic order cancel out while the third order terms, after integration in momenta, are still proportional to the mass. Therefore, in the massless limit $\langle \nabla_{\mu}J^{\mu}_A \rangle_r=0$,  in agreement with the fact that the axial current anomaly obtained from other renormalization prescriptions  $\epsilon^{\mu\nu\alpha \beta}R_{\mu\nu}^{\ \  \lambda \xi}R_{\alpha\beta\lambda\xi}$  vanishes for a FLRW spacetime. In contrast,
the fourth-order adiabatic subtraction terms in the trace of the stress-energy tensor  survive, and, after integration, turn out to be independent of $m$ 
\be  \langle T^{\mu}_{\mu} \rangle_{r}= \frac{1}{240\pi^2} \frac{-4 \dot a^2 \ddot a+9 a \dot a \dddot a+3a \left(\ddot a^2+a \ddddot a\right)}{a^3} \ . \ee
This result should be expressed as a linear combination of the covariant scalars: the Gauss-Bonnet invariant $G$ [which for a FLRW spacetime is given by $G=-2(R_{\mu\nu}R^{\mu\nu} - R^2/3)$], $\Box R$, and $R^2$ (for a FLRW spacetime 
the conformal tensor vanishes identically). We get
$ \langle T^{\mu}_{\mu}\rangle_r = \frac{1}{2880 \pi^2}(\frac{11}{2}G + 6\Box R)$, where
the numerical coefficients for $G$   and $R^2$ coincide exactly with those obtained from other renormalization prescriptions \cite{birrell-davies}, in agreement with the axioms of renormalization in curved spacetime \cite{Waldbook}. The obtained coefficient for $\Box R$ also coincides with the one predicted by other methods. The vanishing of the term proportional to $R^2$ in the trace anomaly can be shown \cite{Parker79} to be a necessary  condition for  the absence of particle creation in a FLRW spacetime in the massless limit, as it is the case for spin-$1/2$ fields.\\

{\it Renormalized stress-energy tensor in de Sitter space.} 
A virtue of the adiabatic  method is its efficiency to perform  computations of renormalized quantities in cosmological backgrounds. Other methods involve very tedious calculations, which are even more complicated for fermions. Using the method developed in this work, and given the modes defining the vacuum, we can  integrate numerically in a straightforward way the renormalized stress-energy tensor of a Dirac field. In the case of de Sitter space, the adiabatic method is also very efficient to perform the integration analytically. The result is
\bea  &\langle T_{\mu\nu} \rangle_r = \frac{1}{960 \pi^2}g_{\mu\nu}(11 H^4 + 130 H^2 m^2 \nonumber \\
& + 120 m^2 (H^2 + m^2)( \log \frac{m}{H}  
 - Re[\psi( -1 + i \frac{m}{H})])) \ , \eea
where $\psi(z)$ is the digamma function.  \\
 

{\it Conclusions.}
In this work we have provided a satisfactory extension of the adiabatic regularization scheme to spin-$1/2$ fields. Our ansatz for the  adiabatic expansion of the fermionic modes differs significantly from the usual WKB-type template used for scalar modes.  We have tested our proposal by the following: i) analyzing particle creation  in de Sitter space,
and ii)  working out the conformal anomaly.  This can be regarded as a nontrivial test of the robustness  of our proposal. As happens for scalar fields, the underlying covariance of the subtraction procedure (based on the covariant notion of adiabatic invariance) makes it a self-consistent renormalization method to deal with spin one-half fields in cosmological backgrounds. We have also showed the power of the method by computing   the renormalized stress-energy tensor of a Dirac field in de Sitter space. Therefore, it opens a new  avenue for many applications  of cosmological relevance.

{\it Acknowledgments:}  J. N-S. would like to thank Leonard Parker for very useful discussions. 
This work is supported  by the  Spanish grant No. FIS2011-29813-C02-02 and  the Consolider Program CPANPHY-1205388.

\end{document}